\documentclass[12pt, fleqn]{article}
\usepackage[cp1251]{inputenc}
\usepackage{latexsym,amsfonts,amssymb}
\usepackage{graphicx}

\usepackage{amsbsy}
\usepackage{amsmath}
\usepackage{epsf}
\usepackage{cite}

\newtheorem{theo}{Theorem}

\newtheorem{remark}{Remark}
\newcommand{\bt}{\begin{theo}}
\newcommand{\et}{\end{theo}}
\newcommand{\bd}{\begin{displaymath}}
\newcommand{\ed}{\end{displaymath}}

\newcommand{\lf}{\left}
\newcommand{\rg}{\right}

\newcommand{\be} {\begin{equation}}
\newcommand{\ee} {\end{equation}}
\newcommand{\ba} {\begin{array}}
\newcommand{\ea} {\end{array}}
\newcommand{\bea}{\begin{eqnarray}}
\newcommand{\eea} {\end{eqnarray}}

\newcommand{\p} {\partial}

\begin{document}

\begin{center}
 {\Large \bf  A mathematical model \\
 \vspace{0.5cm}
   for the coronavirus COVID-19 outbreak   }
\medskip

{\bf Roman Cherniha \footnote{\small  Corresponding author. E-mail: r.m.cherniha@gmail.com}}
  {\bf and  Vasyl' Davydovych \footnote{\small  E-mail:davydovych@imath.kiev.ua }}
 \\
{\it ~Institute of Mathematics,  National Academy
of Sciences  of Ukraine,\\
 3, Tereshchenkivs'ka Street, Kyiv 01004, Ukraine
}\\
 \end{center}

 \begin{abstract}
 A  mathematical model is proposed for quantitative description of the outbreak of novel coronavirus COVID-19 in China. Although the model is relatively simple, the comparison with the public   data   shows that an exact solution  solution of the model  (with the correctly-specified parameters) leads to the results, which are in good agreement with the measured data.
 Prediction of the total number of the COVID-19  cases  is discussed and an example is presented using    the measured data in Austria.
\end{abstract}



\section{Introduction} \label{sec-1}

The outbreak of novel coronavirus called COVID-19 in China has attracted extensive attention of many scientists, in particular mathematicians working in mathematical modeling. The first papers were already published in February and March  2020 \cite{luo,china-19-02-20,shao,tian,roda-michaelLi}.
At the present time, there is an oblivious threat that the COVID-19 outbreak will spread over the world as a pandemic. There were almost 1300000 coronavirus cases up to date  April 6 \cite{meters}.

At the present time, there are many mathematical models used to describe epidemic processes and
they can be found in any book devoted to mathematical models in biology
and medicine (see, e.g., \cite{brauer-12,k-r-2008,mur2,mur2003}  and papers cited therein).
The paper \cite{kermack-1927} is one of the first papers in this direction. The authors created a model based on three ODEs, which nowadays is called the SIR model. There are several  generalizations of the SIR model and the SEIR model \cite{anderson,dietz}, which involves four ODEs, is the most common among them.

Here (Sections~\ref{sec-2} and \ref{sec-3}) we propose a simple model, which was developed using the data from \cite{meters} in the case of the COVID-19 outbreak in China. This case was used because there are obvious indications that this epidemic threat was effectively removed in China. A prediction of the total number of the COVID-19  cases  is discussed and an example is presented using    the measured data in Austria (Section~\ref{sec-4}).

 \section{Mathematical model} \label{sec-2}

 The first nontrivial  biological model used for calculation and the time evolution of the total world
population of people was created in 1838 by
Verhulst \cite{verh-1838}.    His model is usually called the logistic
model  and has the form (in dimensionless
variables)
\[ \frac{dU}{dt} =  U(1-U),
 \quad U(0)=N_0>0, \]
and is the classical example in any textbook on Mathematical
Biology. Its  exact solution  is well known
\be\nonumber U(t) = \frac{N_0e^t}{1+N_0\lf(e^t-1\rg)} \ee
 and depending on the value $N_0$  suggests three different scenarios for
  the population evolution. In particular, the useful curve, the so-called  sigmoid, is obtained if $N_0< 1/2$ (see, e.g., Fig. 1.1 in \cite{ch-dav-book}).

  It can be noted that the data for the total COVID-19 cases in China \cite{meters} can be approximated by   a sigmoid with the correctly-specified parameters.
  Having this in mind, we introduce a smooth function $u(t)$, which  presents the total number of the COVID-19 cases identified up to day $t$ (for any integer number $t$). We assume that the first case (cases) $u_0$ was (were) identified at $t=0.$  Obviously, the function $u(t)$ is non-decreasing. So, we obtain
 \be\label{2-1} \frac{du}{dt} =  u(a-bu),
 \quad u(0)=u_0\geq0 \ee
 where $a$ and $b$ are positive constants.
 One may define $a$  as $a_0S$, where $a_0<1$ is the infection rate  and $S$  is an average number of healthy persons, who was contacted by a fixed infected person. Obviously, each infected person can be in contact only with a limited number of people (usually it is relatives and close friends). The term $bu$  has an opposite meaning  to $a$, because one reflects  the  efforts  $B$, in order to avoid contacts with infected persons and to make other restrictions defined by the government. The coefficient $B$  should increase with growing  $u(t)$. In other words, the government and ordinary people should apply stronger measures in order to stop  growing  $u(t)$, otherwise the control on  the epidemic process will be lost. So, we assume that  $B \approx b_*u^{1+\gamma}$ with $\gamma >0$, therefore the term $b_* u^{1+\gamma}$ (here $b_*>0$)  leading to the equation
 \be\label{2-1*} \frac{du}{dt} =  u(a-b_*u^{\gamma}) \ee
  is derived.  In the case $\gamma =1$, Eq. (\ref{2-1*}) coincides with (\ref{2-1}). We note that the nonlinearity in (\ref{2-1*}) was introduced in \cite{ayala-et-al-73} for describing competition  between species, while the logistic equation in epidemiology occurs naturally and it is  shown  under  some  general assumptions in \cite{brauer-12}.

    During the epidemic process there are two possibilities for the infected persons. A majority, say $w$,  among them will
 recover, while some people, $v$,  will die. Obviously, the equality
  \be\nonumber  u = v +w \ee
  takes place at  any time $t$.  A typical equation for the time evolution of $v$ (see  the last equation in the SIR model)  is
 \be\label{2-2} \frac{dv}{dt} = k(t)u, \quad v(0)=v_0\geq0 \ee
 (a similar equation can be written for $w$ but there is no need to use more equations),
 where $v_0$  is the number of deaths at $t=0$.
 Here the coefficient  $k(t)>0$   reflects the effectiveness of the health system of the country (or a region) in question. From mathematical point of view, this  coefficient should have the asymptotic behavior   $k(t)\to 0$, if $ t \to \infty$, otherwise all infected people will die. In particular, the useful form is $k(t)= k_0\exp(-\alpha t), \, \alpha>0 $.

 \section{Application for the COVID-19 outbreak in China} \label{sec-3}

 The general  solution of Eq. (\ref{2-1})  is well-known, so that Eq. (\ref{2-2}) with the given function $k(t)$
  can be easily integrated. So, setting $k(t)= k_0\exp(-\alpha t), \, \alpha>0 $, we arrive at the exact solution of the model (\ref{2-1}) and (\ref{2-2})
 \be\label{3-1} \ba{l} \medskip  u(t)=\frac{au_0e^{at}}{a+bu_0(e^{at}-1)},
 \\ v(t)= ak_0u_0\int^t_0\frac{e^{(a-\alpha)\tau}}{a+bu_0(e^{a\tau}-1)}\,d\tau+v_0.
 \ea\ee

\begin{remark} The integral in (\ref{3-1}) leads to an expression involving the special function  $LerchPhi(y,c,\nu) \equiv \sum^\infty_{n=0} \frac{y^n}{(\nu+n)^c}$, which  cannot be expressed in terms of elementary functions for arbitrary parameters $\alpha$ and $a$. However, it can be done in some specific cases. For example, one obtains
\[v(t)=\frac{2k_0\sqrt{u_0}}{\sqrt{b(a-bu_0)}} \left(\arctan\left(\frac{\sqrt{bu_0}}{\sqrt{a-bu_0}}\,e^{\frac{at}{2}}\right)-
\arctan\left(\frac{\sqrt{bu_0}}{\sqrt{a-bu_0}}\right)\right)+v_0\]
in the case $2\alpha=a$.
\end{remark}

\begin{figure}[ht]\begin{center}
 \includegraphics[width=9cm]{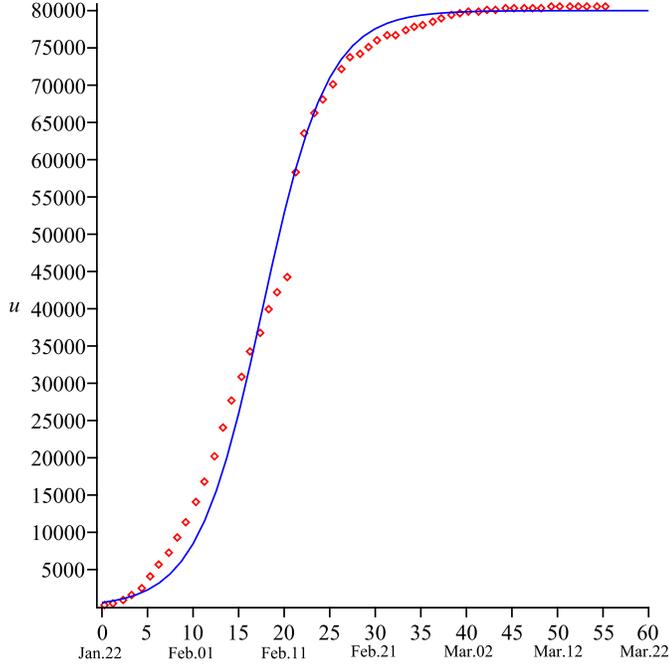}
\end{center}
 \caption{The comparison of the  exact solution $u(t)$ (\ref{3-1})  with $a=0.28, \ b=\frac{7}{2000000}, \ u_0=571$ (blue curve)  and the measured data of the COVID-19 cases (red dots).}\label{f-1}
\end{figure}

 Now we need to specify all the parameters in (\ref{3-1}) using the data for the COVID-19 outbreak in China.
 It follows from \cite{meters} that the earliest well-founded data were fixed on Jan.22, hence we fix this date as $t=0$ and immediately obtain $u_0=571$  and $v_0=17$.
 The parameter $b$  can be found  from the known asymptotic behavior of the function $u(t)$ in (\ref{3-1}) and information from \cite{meters}, therefore $b \approx\frac{a}{80000}$.
 The plausible interval for  parameter $a$ can be estimated
 by using option  `animation' in MAPLE in order to fit plot of the function $u(t)$ to the measured  data after Jan.22.
 So, we have numerically proved  that  $a  \in [0.25, 0.30]$.

\begin{figure}[ht]\begin{center}
  \includegraphics[width=9cm]{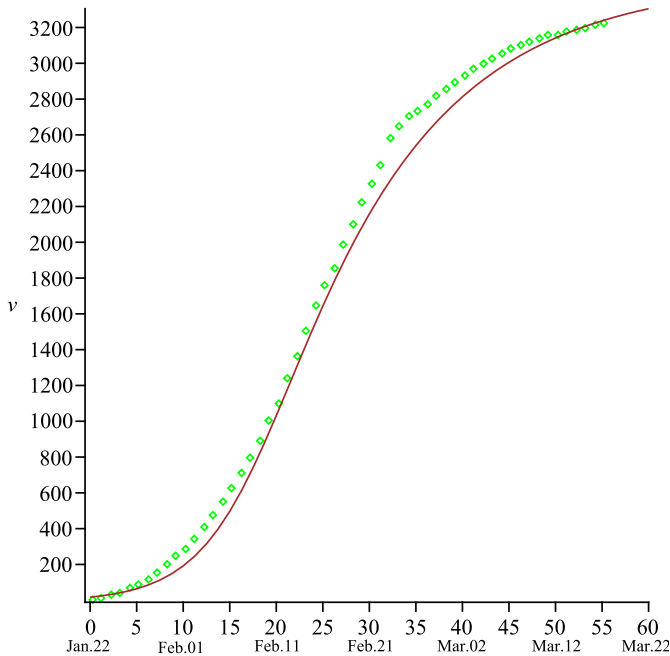}
\end{center}
 \caption{The comparison of the exact solution $v(t)$ (\ref{3-1}) with  $a=0.28, \ b=\frac{7}{2000000}, \ u_0=571, \ k_0=0.0094, \ \alpha=0.07, \ v_0=17$ (brown curve) and the measured data the total  number of  deaths (green dots).}\label{f-2}
\end{figure}

 Because the function $v(t)$ should be monotonic non-decreasing function (we remind that it is the number of total deaths), we conclude that $a>\alpha$. It was identified that a good choice is $4\alpha =a$. Finally,  the coefficient
 $k_0$  was found from the formula
\be\nonumber v\big|_{t=T}\equiv571k_0\int^{T}_0\frac{e^{0.21\tau}}{1+\frac{571}{80000}(e^{0.28\tau}-1)}\,d\tau+17=V\ee (here $V$ is the number of total deaths in the time $t=T$ presented in \cite{meters}) for the fixed value of $a=0.28$.
The value of the coefficient $k_0$ is slowly varied  from $0.0092385$ to $0.0096878$ if $T$ is changed from $65$ to $45$, respectively. So, the value $k_0=0.0094$ was chosen.

 Fig.~\ref{f-1} and Fig.~\ref{f-2}  present the comparison of the  results  obtained from the model (\ref{2-1}) and (\ref{2-2}) (with  the parameters specified  above)  and the measured  data for the COVID-19 outbreak in China \cite{meters}.
 One may note that there is a good agreement  between  the total  number of  the COVID-19 cases  and that predicted by our model. Of course, one may claim that exactness is not sufficiently good in the interval $ t \in [10,25]$ in Fig.~\ref{f-1}. However, we assume that either the method of measurement of the COVID-19 cases was corrected, or an unpredictable spike of such cases occurred around date $t=25$ (there is a jump from 44653 cases on Feb.11 to 58761 cases on Feb.12 \cite{meters}).

 The comparison  between the total  number of  deaths  and that predicted by our model shows that exactness is  sufficiently good for any time (see Fig.~\ref{f-2}). One may also note that the function $v(t)$ is still increasing beyond the time $t=60$. Such behavior reflects the real situation in the epidemic process, namely: some   people will die even in absence of new COVID-19 cases because they were infected earlier. So, the final number of  total deaths will be fixed later than that of the COVID-19 cases.

\section{Discussion} \label{sec-4}
In this work, a mathematical model is proposed for quantitative description of the outbreak of novel coronavirus COVID-19 in China. Although the moodel is relatively simple, the comparison with the  data  listed in  \cite{meters} shows that the analytical solution of the model  (with the correctly-specified parameters) leads to the results, which are in good agreement with the measured data.

Some well-known recommendations naturally follow from the model. It follows from the exact solution  (\ref{3-1}) that one needs to reduce  the coefficient $a= a_0S$ as much as possibly. It means that the number of contacts $S$  should be minimized. On the other hand, the government should make more efforts (to close shops, restaurants, to restrict transport traffic etc.)  in order to increase the function $B(u)$. These efforts should increase with growing of the total  number of  the COVID-19 cases.
The government restrictions can be stopped only under condition that
 that the number of new  COVID-19 cases per day already began  to decrease from day to day. It means mathematically that the second order derivative of the function $u(t)$  takes negative values.
In order to find the so-called critical  number $u^*$, we analyze the function $u(t)$ from (\ref{3-1}). Calculating the second order derivative, one obtains
\be\nonumber
u'' = \frac{u_0a^3\lf(a-bu_0\rg)e^{at}\lf(a-bu_0-bu_0e^{at}\rg)}
{\lf(a-bu_0+bu_0e^{at}\rg)^3}.
\ee
Solving the algebraic equation $u''=0$ with respect to the time, we arrive at \be\nonumber t^* = \frac 1a \ln\lf(\frac{a}{bu_0}-1\rg), \ee
hence $u^*= u(t^*).$
On the other hand, formula $u''=0$ allows to find the parameter $b$ provided the time $t^*$  is known from the measured data. Assuming that $a$ is known one calculates
 \be\label{4-4} b= \frac{a}{u_0\lf(e^{at^*} + 1\rg)}. \ee
Taking into account interpretation of the parameters, we believe that the parameter $a$ varies not so much as $b$  and can be specified (at least estimated with a sufficient exactness) as follows.
Obviously, the total number of the COVID-19 cases in the initial period of epidemic process can be approximated as   $u(t)\approx u_0 e^{at}$ (see $u(t)$ in (\ref{3-1}) for small time). So, having the measured data  in the initial period, we may specify the parameter $a$.
 It means that our model can allow to predict  the total numbers of the COVID-19 cases if the data for  $t^*$  and $u^*$  are known.

\begin{figure}[ht]\begin{center}
 \includegraphics[width=8.5cm]{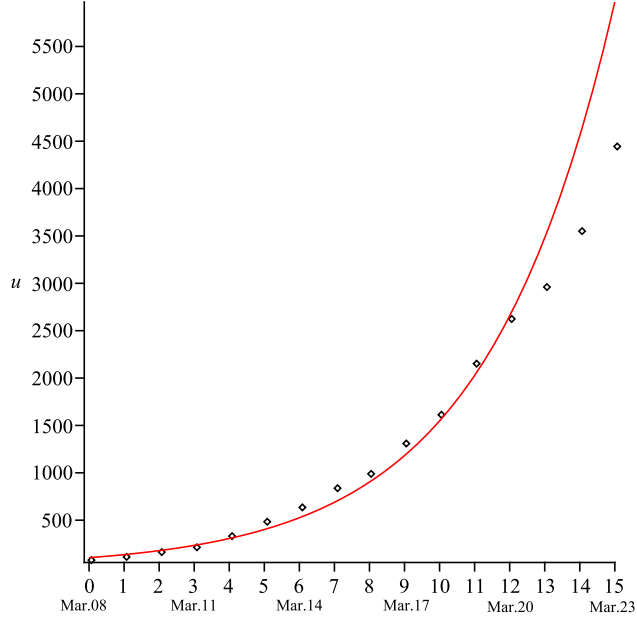}
\end{center}
 \caption{An exponential growing of the total number of the COVID-19 cases in Austria at the initial stage of the coronavirus spread (red curve)  and the measured data (black dots)} \label{f-3}
\end{figure}

Let us consider, an example. It can be noted from the public data  \cite{meters}
that the COVID-19 outbreak in Austria had  the maximum number of new daily cases on March 26. So, $u^*= 6909$. If we fix March 8 as the initial point  $t=0$, then $t^*= 18$   and  $u_0=104$  (we think that there are essential errors in measuring  at the very beginning of  the epidemic process, so that it is unreasonable to start from very small numbers of  $u_0$). Now we make approximation of the measured COVID-19 cases using the formula $u(t)\approx 104 e^{at}$ during the first 15 days. It turns out that  the parameter $a=0.27$  provides very good approximation during the first 12 days (see Fig.~\ref{f-3}). So, using formula (\ref{4-4})
we define $b=0.000020$. Now  we may predict that the total number  the COVID-19 cases  in Austria  should be $u_{max}=a/b \approx 13 500$. Taking into account that this number was  calculated under some assumptions,  the real number can be  larger. For exmple, if one takes March 10 as the initial point  $t=0$  then  $u_{max}\approx 13 800$. We estimate the maximum error in 10 percent.

\begin{figure}[ht]\begin{center}
 \includegraphics[width=8.5cm]{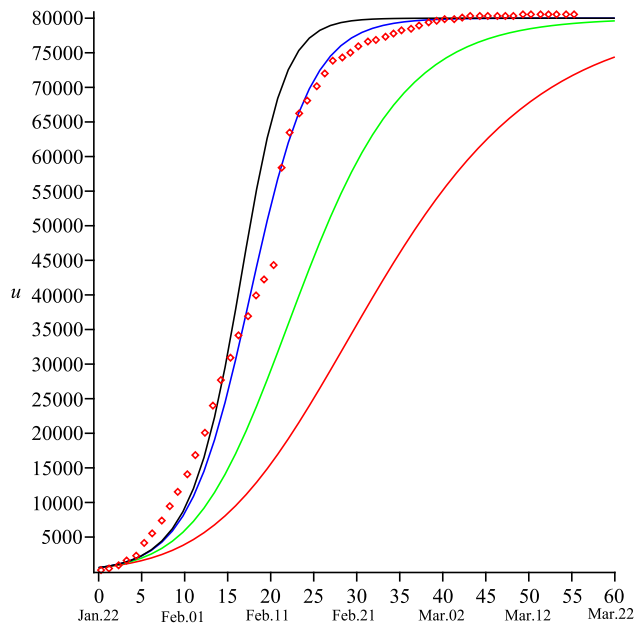}
\end{center}
 \caption{The solution $u(t)$ of Eq. (\ref{2-1*})  for  $\gamma=0.3$ (red curve), $\gamma=0.5$ (green curve), $\gamma=1$ (blue curve), $\gamma=1.5$ (black curve) and  the measured data of the COVID-19 cases in China(red dots).}\label{f-4}
\end{figure}

It should be noted that the parameter $\gamma$  plays essential role if one uses  Eq. (\ref{2-1*}) instead of Eq. (\ref{2-1}). In order to highlight this, we present exact solutions of  Eq. (\ref{2-1*}) with different values of $\gamma$ in Fig.~\ref{f-4} (all other parameters are the same as in Fig.~\ref{f-1}). One may see that $\gamma=1$ is  a good  choice in the case of China. On the other hand, taking into account the known  data   \cite{meters}, we conclude that $\gamma<1$ in the case of  S. Korea.

Obviously, the model cannot be thought as such that is applicable for the  COVID-19  outbreak in each country. For example, the outbreak in China was mostly localized in the province Hubei. The size and population of this province are very small comparing with total those of China. A similar situation is in USA, where  two states, New-York and New-Jersey are affected by the coronavirus much more than other states (up to date April 5, 2020).

 On the other hand, if we take the epidemic process in  Italy  then one notes that the size and population of Northern Italy (8 provinces, Lombardia is the largest among them)  are comparable with those all of Italy. So, we propose that the space distribution of the infected population should be taken into account in such cases.
The simplest generalization of the basic equations of our model are
\be\label{4-1}\begin{array}{l} \medskip  \frac{\p u}{\p t} = d_1 \Delta u+u(a- bu^\gamma),  \\
\frac{\p v}{\p t} = d_2\Delta u +k(t)u, \end{array}\ee
 where $\Delta $ is
the Laplace operator, $d_1$ and  $d_2$  are diffusivities,
 the functions $u(t,x,y)$  and $v(t,x,y)$ are analogs of $u(t)$  and $v(t)$.
 Of course, the generalized model  based on the system of  (\ref{4-1}) and relevant boundary conditions (for example, zero flux conditions at the boundary) is much more complicated  problem  and cannot be solved analytically  in contrast to the model (\ref{2-1}) and (\ref{2-2}).  Here we only note that the first equation in  (\ref{4-1}) with $\gamma=1$ is the classical Fisher equation \cite{fi}, which was extensively studied in many works (see, e.g., the monographs \cite{mur2,ch-se-pl-book} and papers cited therein).

\end{document}